\documentstyle[12pt]{article}
\begin{document}
\parskip=5pt plus 1pt minus 1pt

\begin{flushright}
{\bf LMU-10/95}\\
{April, 1995}
\end{flushright}

\vspace{0.2cm}
\begin{center}
{\Large\bf Comments on the Characteristic Measurables \\ of the Quark Mixing
Matrix}
\end{center}

\vspace{0.7cm}
\begin{center}
{\bf Zhi-zhong XING} \footnote{E-mail address:
Xing@hep.physik.uni-muenchen.de}
\end{center}
\begin{center}
{\sl Sektion Physik, Theoretische Physik, Universit$\ddot{a}$t
M$\ddot{u}$nchen, \\
Theresienstrasse 37, D-80333 M$\ddot{u}$nchen, Germany}
\end{center}

\vspace{3cm}
\begin{abstract}
Within the standard electroweak model we point out that the $3\times 3$ matrix
of quark
mixing is characterized by three universal (rephasing-invariant) quantities:
one of them
for $CP$ violation and the other two for off-diagonal asymmetries. Unitarity of
the quark
mixing matrix can in principle be tested through a variety of measurements
which are
irrelevant to the existence of $CP$-violating effects.
\end{abstract}

\vspace{4cm}
\begin{center}
\end{center}

\newpage

In the standard electroweak model, the $3\times 3$ Cabibbo-Kobayashi-Maskawa
(CKM) matrix
$V$ provides a natural description of quark mixing and $CP$ violation [1,2].
Unitarity is
the only but powerful constraint, imposed by the model itself, on $V$. This
restriction is
commonly expressed as two sets of orthogonality-plus-normalization conditions:
$$
\sum_{k=1}^{3}V_{ik}V^{*}_{jk} \; =\; \delta_{ij} \; , \;\;\;\;\;\;\;\;\;
\sum_{i=1}^{3}V_{ij}V^{*}_{ik} \; =\; \delta_{jk} \; ,
\eqno{(1)}
$$
where $i,j,k=1,2,3$, running over the up-type quarks ($u,c$ and $t$) and
the down-type quarks ($d,s$ and $b$). In the complex plane the six
orthogonality relations
given above correspond to six triangles (see Fig. 1), the so-called unitarity
triangles [3].
Confronting these unitarity requirements with the existing and forthcoming
experimental data
may serve for a stringent test of the standard model.

\vspace{0.3cm}

By use of the unitarity conditions in Eq. (1), one can parametrize the CKM
matrix in various
ways. Several popular parametrizations [2,4,5], including the ``standard'' one
[6], are given
in terms of three Euler angles and one $CP$-violating phase. It is interesting
to note the
fact that knowledge of only the magnitudes of four independent $V_{ij}$ is
sufficient
to determine all phase information and construct the entire matrix $V$ [7]. On
the other hand,
four independent angles (inner or outer) of the six unitarity triangles, once
they are
measured from the $CP$ asymmetries in weak $B$-meson decays, can also determine
the whole
quark mixing matrix [8,9].

\vspace{0.3cm}

In this note we shall point out that the CKM matrix $V$ is in fact
characterized by three
universal (rephasing-invariant) quantities: one of them for $CP$ violation and
the other two
for the off-diagonal asymmetries of $V$. These measurables have interesting
relations with the
six unitarity triangles. We briefly comment on some possibilities to test
unitarity of the
quark mixing matrix through the accessible measurements at present or in the
near future.
We stress the point that the Kobayashi-Maskawa mechanism of $CP$ violation [2]
can experimentally
be checked even in the absence of direct observation of $CP$-violating signals.

\vspace{0.3cm}

It is well known that unitarity of the CKM matrix leads to a universal and
rephasing-invariant
measure of $CP$ violation for quark weak interactions [10,11], the so-called
Jarlskog parameter
$J$ [10]:
$$
{\rm Im}\left (V_{il}V_{jm}V^{*}_{im}V^{*}_{jl}\right ) \; =\;
J\sum_{k,n=1}^{3}\epsilon_{ijk}
\epsilon_{lmn} \; .
\eqno{(2)}
$$
One can show that all the six unitarity triangles have the same area $J/2$,
although
their shapes are quite different (see Fig. 1 for illustration). Here the
interesting point is
that $J^{2}$ can be simply expressed in terms of three sides of each triangle:
$$
J^{2} \; = \; 4P_{i}\prod^{3}_{l=1}\left (P_{i}-|V_{jl}V^{*}_{kl}|\right ) \; =
\;
4Q_{i}\prod^{3}_{l=1}\left (Q_{i}-|V_{lj}V^{*}_{lk}|\right ) \; ,
\eqno{(3a)}
$$
where the subscripts $i,j$ and $k$ (=1,2,3) must be co-cyclic for the up- or
down-type quarks,
$P_{i}$ and $Q_{i}$ are given by
$$
P_{i} \; =\; \frac{1}{2}\sum^{3}_{l=1}|V_{jl}V^{*}_{kl}| \; , ~~~~~~~
Q_{i} \; =\; \frac{1}{2}\sum^{3}_{l=1}|V_{lj}V^{*}_{lk}| \; .
\eqno{(3b)}
$$
Clearly $P_{i}$ and $Q_{i}$ correspond to the unitarity triangles $[u], [c],
[t]$ and
$[d], [s], [b]$ in Fig. 1, respectively. This result implies that one can in
principle obtain
the information about $CP$ violation only from the sides of the unitarity
triangles. In this
way we are able to experimentally check unitarity of the quark mixing matrix
and the
Kobayashi-Maskawa picture of $CP$ violation, without the help of direct
measurements of any
$CP$-violating signal.

\vspace{0.3cm}

The off-diagonal asymmetries of the CKM matrix are another two universal and
characteristic
quantities for quark mixing. They are denoted by $Z_{1}$ about the
$V_{11}-V_{22}-V_{33}$
axis [12] and $Z_{2}$ about the $V_{13}-V_{22}-V_{31}$ axis:
$$
|V_{ij}|^{2}-|V_{ji}|^{2} \; =\; Z_{1}\sum^{3}_{k=1}\epsilon_{ijk} \; , ~~~~~~
|\hat{V}_{ij}|^{2}-|\hat{V}_{ji}|^{2} \; =\; Z_{2}\sum^{3}_{k=1}\epsilon_{ijk}
\; ,
\eqno{(4a)}
$$
where the matrix $\hat{V}$ is obtained from $V$ through the following rotation:
$$
\hat{V} \; =\; VR \; , ~~~~~~~ R \; =\; \left (
\begin{array}{ccc}
0	& 0	& 1 \\
0	& 1	& 0 \\
1	& 0	& 0
\end{array}
\right ) \; .
\eqno{(4b)}
$$
Certainly the axis $\hat{V}_{11}-\hat{V}_{22}-\hat{V}_{33}$ of $\hat{V}$ is
equivalent to
the axis $V_{13}-V_{22}-V_{31}$ of $V$. The above result can be shown easily by
use of the
normalization conditions of unitarity given in Eq. (1). Note that the asymmetry
parameters
$Z_{1,2}$ are independent of each other, and they are independent of the
$CP$-violating
parameter $J$. Although a little attention was paid to $Z_{1}$ in the
literature [12,13],
$Z_{2}$ has been ignored. We shall see later on that $Z_{2}>>Z_{1}$, and both
of them play
interesting roles in testing unitarity of the quark mixing matrix.

\vspace{0.3cm}

Explicitly One can express $J$ and $Z_{1,2}$ with any set of parameters
suggested in
Refs. [2,4-6]. To illustrate these characteristic measurables in a simple and
instructive way,
here we use the Wolfenstein parameters [4]. Note that a self-consistent
calculation of
$Z_{1,2}$ can only be carried out on the basis of the modified Wolfenstein
parametrization [12]
\footnote{In terms of the Wolfenstein parameters, Kobayashi has presented a
parametrization
of the CKM matrix with exact unitarity [14]. An incomplete modification of the
Wolfenstein
parametrization, where the imaginary parts of $V_{21}$ and $V_{32}$ are
corrected up to the
accuracy of $O(\lambda^{5})$, was given by Buras {\it et al} for their own
purpose in Ref. [15].
About seven years before, Branco and Lavoura have parametrized the CKM matrix
up to $O(\lambda^{8})$
by taking $V_{12}=\lambda, V_{23}=A\lambda^{2}$ and
$V_{13}=A\mu\lambda^{3}e^{{\rm i}\phi}$ [16].},
in which unitarity is kept up to the accuracy of $O(\lambda^{6})$. In terms of
$\lambda, A, \rho$
and $\eta$ [12,14], we obtain
$$
Z_{1} \; \approx \; A^{2}\lambda^{6}(1-2\rho) \; , ~~~~~~~
Z_{2} \; \approx \; \lambda^{2}\left (1-A^{2}\lambda^{2}\right ) \; , ~~~~~~~
J \; \approx \; A^{2}\lambda^{6}\eta \; .
\eqno{(5)}
$$
It is clear that $Z_{2}>>Z_{1}$ and $Z_{1}\sim J$. Both $Z_{1}$ and $Z_{2}$ are
independent of
$\eta$, a parameter necessary for signalling $CP$ violation. Considering the
values of
$\lambda, A, \rho$ and $\eta$ extracted from experiments [17], we find
$Z_{2}/Z_{1}\geq 400$,
$Z_{1}\sim 10^{-5} - 10^{-4}$ and $J\sim 10^{-5}$. The possibility of
$Z_{1}\approx 0$, which
requires $\rho\approx 0.5$, is only allowed on the margin of the current data
[12].

\vspace{0.3cm}

We are in a position to give a few comments on the measure of $CP$ violation,
the unitarity
triangles and the possibilities to test unitarity of the CKM matrix:

\vspace{0.3cm}

(a) Although all measurables of $CP$ violation are proportional to $J$ in the
standard model,
their magnitudes are more sensitively related to the angles of the unitarity
triangles
(see, e.g., Refs. [8,9]). This is why different weak processes of quarks may
have different sizes of
$CP$-violating effects. In this sense whether the value of $J$ is maximal or
not has less
physical significance than the maximal violation of $P$ (parity) symmetry does.

\vspace{0.3cm}

(b) In general the six unitarity triangles have nine different inner (or outer)
angles, although
they have eighteen different sides (see Fig. 1). If $Z_{1}=0$ holds, one can
find three
equivalence relations among the six triangles:
$$
[u] \; \cong \; [d] \; , ~~~~~~~
[c] \; \cong \; [s] \; , ~~~~~~~
[t] \; \cong \; [b] \; .
\eqno{(6)}
$$
In this case the six unitarity triangles have six different inner (or outer)
angles and nine
different sides. As a consequence the CKM matrix can be parametrized by use of
three independent
quantities. If we further assume $Z_{2}=Z_{1}=0$, then two sides of the
triangle $[c]$ or $[s]$
would become equal (i.e., $|V_{31}V^{*}_{11}|=|V_{33}V^{*}_{13}|$ etc). Today
the possibility of
$Z_{2}=0$ has been ruled out absolutely, while that of $Z_{1}=0$ is still in
marginal agreement
with the experimental restriction \footnote{The current data have given
$|V_{12}|=0.2205 \pm
0.0018$ and $|V_{21}|=0.204 \pm 0.017$ [6], which implies a dominant
possibility of $Z_{1}>0$.}.

\vspace{0.3cm}

(c) If $Z_{1}>0$ is really true, then one can find the following hierarchical
relation among the nine
matrix elements:
$$
|V_{33}| \; > \; |V_{11}| \; > \; |V_{22}| \; >> \; |V_{12}| \; > \; |V_{21}|
\; >> \;
|V_{23}| \; > \; |V_{32}| \; >> \; |V_{31}| \; > \; |V_{13}| \; .
\eqno{(7)}
$$
This interesting result reflects the unitarity of the CKM matrix in an indirect
way.

\vspace{0.3cm}

(d) It is instructive to express the nine inner angles of the unitarity
triangles in terms of
the Wolfenstein parameters. To lowest order approximation, we obtain the
following results
(see Fig. 1):
$$
\tan (\angle 1) \; \approx \; \tan (\angle 4) \; \approx \; - \tan (\angle 3)
\; \approx \; \frac{\eta}
{1-\rho} \; ,
\eqno{(8a)}
$$
$$
\tan (\angle 6) \; \approx \; \tan (\angle 7) \; \approx \; -\tan (\angle 9) \;
\approx \; \frac{\eta}
{\rho} \; ,
\eqno{(8b)}
$$
and
$$
\tan (\angle 2) \; \approx \; \lambda^{2}\eta \; , ~~~~~~~~~
\tan (\angle 8) \; \approx \; A^{2}\lambda^{4}\eta \; , ~~~~~~~~~
\tan (\angle 5) \; \approx \; \frac{\eta}{\rho (\rho -1)+\eta^{2}} \; .
\eqno{(8c)}
$$
Conventionally one uses $\alpha =\angle 5$, $\beta =\angle 1$ and $\gamma
=\angle 7$ to denote
the three angles of unitarity triangle $[s]$, which will be overdetermined at
$B$-meson factories.
We expect that the approximate relations given in Eq. (7) can be tested in
various experiments
of $CP$ violation and $B$ physics in the near future [18].

\vspace{0.3cm}

(e) The magnitude of $J$ is now determinable from the unitarity triangle $[t]$,
whose three
sides have all been measured in weak decays of the relevant quarks (or from
deep
inelastic neutrino scattering [6]). The off-diagonal asymmetries of $V$ can be
directly
determined from the experimental data on $V_{12}, V_{21}$ and $V_{23}$, i.e.,
$Z_{1}=|V_{12}|^{2}-|V_{21}|^{2}$ and $Z_{2}=|V_{12}|^{2}-|V_{23}|^{2}$. With
the help of
precise information about $B^{0}_{d}-\bar{B}^{0}_{d}$ mixing, one is able to
determine
$|V_{33}V^{*}_{31}|$ and then to establish the unitarity triangle $[s]$. A
comparison
between the areas of $[t]$ and $[s]$ may serve to confirm the unitarity
conditions of the
CKM matrix. Similarly the forthcoming measurement of
$B^{0}_{s}-\bar{B}^{0}_{s}$ mixing
will determine $|V_{32}V^{*}_{33}|$ and construct the triangle $[d]$. It might
be a long
run to directly measure $|V_{31}|$ and $|V_{32}|$ from the production or decay
processes of
the top quark.

\vspace{0.3cm}

(f) The normalization relations of unitarity (see Eq. (1)) can be well checked
after
more precise determination of $|V_{13}|$ from charmless $B$-meson decays and
measurements of $|V_{33}|$ from the top-quark lifetime. To test unitarity of
the CKM
matrix up to $O(\lambda^{6})$, of course, much effort is needed to make in
order to
improve the accuracy of the six elements in the first two rows of $V$.
In practice observation of $CP$ violation in $B$-meson decays will provide a
good
chance to judge the Kobayashi-Maskawa mechanism of $CP$ violation as well as
the
six orthogonality conditions of unitarity. Discussions about violation of
unitarity
of the $3\times 3$ CKM matrix, e.g., in the presence of the fourth-family
quarks or
an exotic charge $-1/3$ quark, would be beyond the scope of this note [19].

\vspace{0.3cm}

In summary, we have pointed out that the $3\times 3$ matrix of quark mixing is
characterized
by three universal observables: the measure of $CP$ violation $J$ and the
off-diagonal
asymmetries $Z_{1,2}$. These three parameters have interesting relations with
the
six unitarity triangles. The Kobayashi-Maskawa picture of $CP$ violation can in
principle
be examined through a variety of measurements irrelevant to the existence of
$CP$-violating
effects. A complete test of unitarity of the CKM matrix (to a good degree of
accuracy)
is accessible in the near future.

\vspace{0.5cm}

The author would like to thank Professor H. Fritzsch for his warm hospitality
and constant
encouragement. He is also grateful to Professors G. C. Branco and D. D. Wu for
useful
discussions, and to Professor R. E. Shrock for helpful comments. This work was
financially
supported by the Alexander von Humboldt Foundation of Germany.

\normalsize

\newpage

\begin{figure}
\begin{picture}(400,350)
\put(70,300){\line(1,0){130}}
\put(120,310){\makebox(0,0){${\bf V}_{cb}{\bf V}^{*}_{tb}$}}
\put(70,300){\line(2,-5){12}}
\put(51,285){\makebox(0,0){${\bf V}_{cd}{\bf V}^{*}_{td}$}}
\put(200,300){\line(-4,-1){118}}
\put(137,268){\makebox(0,0){${\bf V}_{cs}{\bf V}^{*}_{ts}$}}

\put(80,294){\makebox(0,0){\scriptsize\bf 1}}
\put(150,294){\makebox(0,0){\scriptsize\bf 2}}
\put(87,278){\makebox(0,0){\scriptsize\bf 3}}
\put(130,235){\makebox(0,0){$[u]$}}

\put(300,300){\line(1,0){112}}
\put(370,310){\makebox(0,0){${\bf V}_{cs}{\bf V}^{*}_{cb}$}}
\put(300,300){\line(4,-1){117}}
\put(350,272){\makebox(0,0){${\bf V}_{ts}{\bf V}^{*}_{tb}$}}
\put(412,300){\line(1,-6){4.9}}
\put(440,282){\makebox(0,0){${\bf V}_{us}{\bf V}^{*}_{ub}$}}

\put(350,294){\makebox(0,0){\scriptsize\bf 2}}
\put(405,293){\makebox(0,0){\scriptsize\bf 9}}
\put(408,280){\makebox(0,0){\scriptsize\bf 6}}
\put(365,235){\makebox(0,0){$[d]$}}
\end{picture}

\begin{picture}(400,160)
\put(85,300){\line(1,0){90}}
\put(127,310){\makebox(0,0){${\bf V}_{td}{\bf V}^{*}_{ud}$}}
\put(85,300){\line(1,-3){13}}
\put(68,278){\makebox(0,0){${\bf V}_{ts}{\bf V}^{*}_{us}$}}
\put(175,300){\line(-2,-1){77}}
\put(155,268){\makebox(0,0){${\bf V}_{tb}{\bf V}^{*}_{ub}$}}

\put(95,294){\makebox(0,0){\scriptsize\bf 4}}
\put(146,294){\makebox(0,0){\scriptsize\bf 5}}
\put(102.5,272.5){\makebox(0,0){\scriptsize\bf 6}}
\put(130,235){\makebox(0,0){$[c]$}}

\put(315,300){\line(1,0){85}}
\put(360,310){\makebox(0,0){${\bf V}_{ub}{\bf V}^{*}_{ud}$}}
\put(315,300){\line(2,-1){73}}
\put(338,270){\makebox(0,0){${\bf V}_{tb}{\bf V}^{*}_{td}$}}
\put(400,300){\line(-1,-3){12}}
\put(420,280){\makebox(0,0){${\bf V}_{cb}{\bf V}^{*}_{cd}$}}

\put(343,294){\makebox(0,0){\scriptsize\bf 5}}
\put(390,293.5){\makebox(0,0){\scriptsize\bf 7}}
\put(385,273){\makebox(0,0){\scriptsize\bf 1}}
\put(365,235){\makebox(0,0){$[s]$}}
\end{picture}

\begin{picture}(400,160)
\put(70,300){\line(1,0){150}}
\put(130,310){\makebox(0,0){${\bf V}_{ud}{\bf V}^{*}_{cd}$}}
\put(70,300){\line(1,-4){6}}
\put(49,286){\makebox(0,0){${\bf V}_{ub}{\bf V}^{*}_{cb}$}}
\put(220,300){\line(-6,-1){143.7}}
\put(157,274){\makebox(0,0){${\bf V}_{us}{\bf V}^{*}_{cs}$}}

\put(79,294){\makebox(0,0){\scriptsize\bf 7}}
\put(150,294){\makebox(0,0){\scriptsize\bf 8}}
\put(81,283){\makebox(0,0){\scriptsize\bf 9}}
\put(130,243){\makebox(0,0){$[t]$}}

\put(280,300){\line(1,0){130}}
\put(355,310){\makebox(0,0){${\bf V}_{cd}{\bf V}^{*}_{cs}$}}
\put(280,300){\line(6,-1){134}}
\put(330,277){\makebox(0,0){${\bf V}_{ud}{\bf V}^{*}_{us}$}}
\put(410,300){\line(1,-5){4.5}}
\put(436,286){\makebox(0,0){${\bf V}_{td}{\bf V}^{*}_{ts}$}}

\put(350,294.5){\makebox(0,0){\scriptsize\bf 8}}
\put(404,294.5){\makebox(0,0){\scriptsize\bf 3}}
\put(407,284.5){\makebox(0,0){\scriptsize\bf 4}}
\put(365,243){\makebox(0,0){$[b]$}}
\end{picture}
\vspace{-6.5cm}
\caption{The unitarity triangles of the CKM matrix in the complex plane.
Each triangle is named in terms of the quark flavor that does not manifest in
its three sides.
Note that the six triangles have the same area, and they only have nine
different inner angles
(versus eighteen different sides).}
\end{figure}
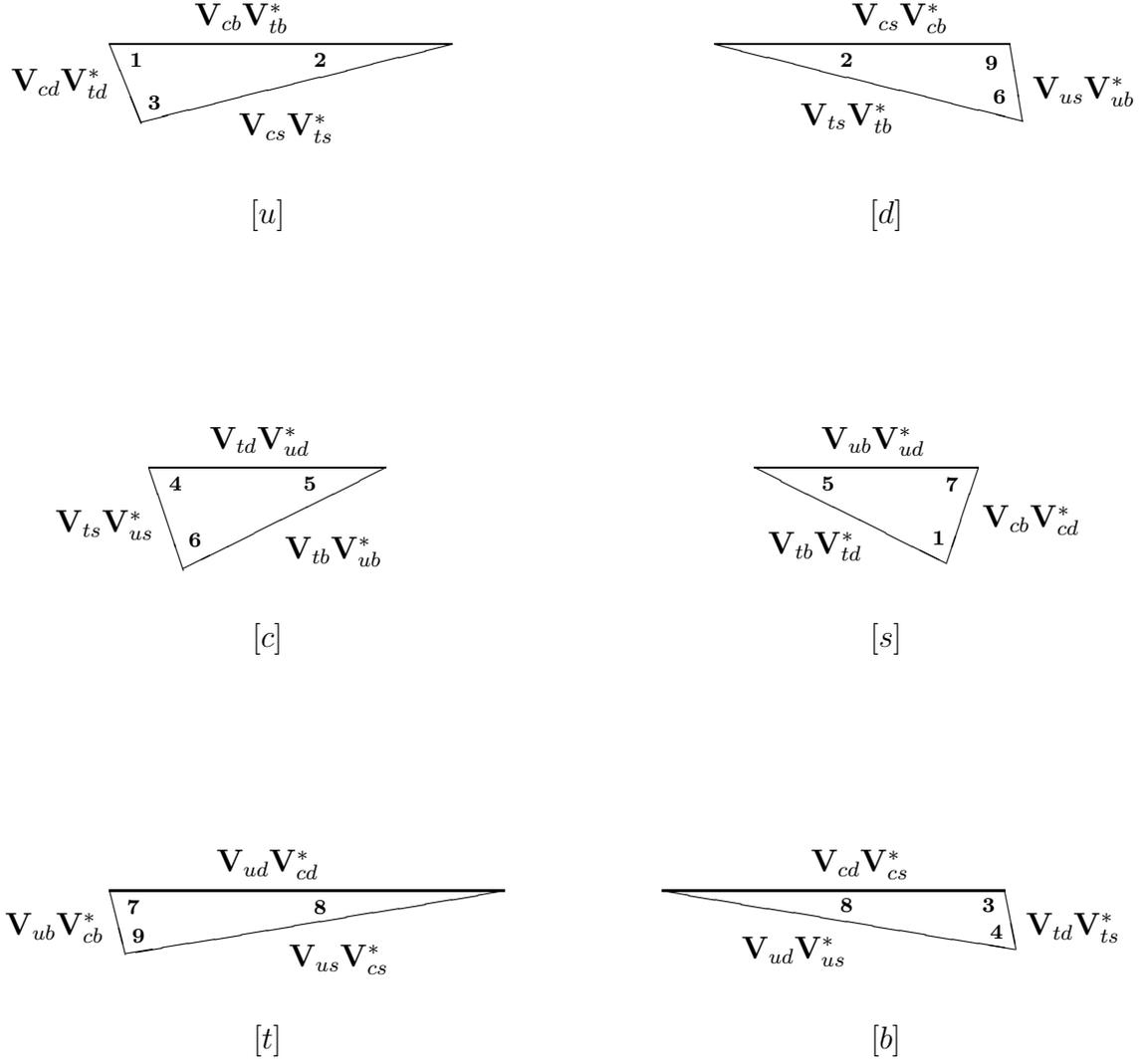

\end{document}